# Exceeding 400% tunnel magnetoresistance at room temperature in epitaxial Fe/MgO/Fe(001) spin-valve-type magnetic tunnel junctions

Thomas Scheike, Qingyi Xiang, Zhenchao Wen, Hiroaki Sukegawa,[*] Tadakatsu Ohkubo, Kazuhiro Hono, and Seiji Mitani

*National Institute for Materials Science (NIMS), Tsukuba 305-0047, Japan*

## Abstract

Giant tunnel magnetoresistance (TMR) ratios of 417% at room temperature (RT) and 914% at 3 K were demonstrated in epitaxial Fe/MgO/Fe(001) exchanged-biased spin-valve magnetic tunnel junctions (MTJs) by tuning growth conditions for each layer, combining sputter deposition for the Fe layers, electron-beam evaporation of the MgO barrier, and barrier interface tuning. Clear TMR oscillation as a function of the MgO thickness with a large peak-to-valley difference of ~80% was observed when the layers were grown on a highly (001)-oriented Cr buffer layer. Specific features of the observed MTJs are symmetric differential conductance (*dI*/*dV*) spectra for the bias polarity and plateau-like deep local minima in *dI*/*dV* (parallel configuration) at |*V*| = 0.2~0.5 V. At 3K, fine structures with two dips emerge in the plateau-like *dI*/*dV*, reflecting highly coherent tunneling through the Fe/MgO/Fe. We also observed a 496% TMR ratio at RT by a 2.24-nm-thick-CoFe insertion at the bottom-Fe/MgO interface.

*Author to whom correspondence should be addressed: sukegawa.hiroaki@nims.go.jp

After the discovery of the "giant tunnel magnetoresistance effect" through the spin-dependent, coherent-tunneling mechanism in magnetic tunnel junctions (MTJs) using a crystalized MgO barrier,[1–5] a tunnel magnetoresistance (TMR) ratio has significantly improved compared with conventional MTJs with amorphous-based barriers, such as $AlO_x$.[6] The large TMR ratios at room temperature (RT) over 100% substantially boosted spintronic applications including read-heads of hard disk drives, magnetoresistive random access memories (MRAMs), and high sensitivity magnetic-field sensors. Many studies based on MTJs with an MgO barrier focused on enhancing the TMR ratio to improve the application range of MTJs. The largest TMR ratio of 604% at RT was demonstrated by Ikeda *et al*. using a $Ta/Co_{20}Fe_{60}B_{20}/MgO/Co_{20}Fe_{60}B_{20}/Ta$ polycrystalline pseudo-spin-valve structure.[7]

However, to create innovative applications using MTJ based technology such as high capacity MRAMs toward storage-class memories,[8] magnetic logic circuits,[9] and brain-morphic devices,[10] achievements of much larger TMR ratios, e.g., >1,000% at RT, are indispensable.[11–13] There is a large discrepancy in TMR ratios between the theoretical limit and experiments. In the widely studied Fe/MgO/Fe(001) structure, first-principles calculations suggested a TMR ratio of over several thousand percent.[2,14–18] However, reported experimental values have barely exceeded 200% at RT. Thus, it is essential to clarify the origin of the discrepancy toward a substantially large experimental TMR ratio. The experimental TMR ratios are extremely sensitive to the state of the Fe/MgO interfaces since it is difficult to obtain sharp and defect-free interfaces owing to the presence of lattice mismatch between Fe and MgO (~3%) and the formation of an undesirable oxide layer and sometimes oxygen vacancies at the interface.[4,19,20] Many misfit dislocations are introduced at the interface, making it challenging to obtain the same quality at the upper and lower MgO interfaces. Significantly asymmetric current–voltage (*I–V*) characteristics are commonly observed in Fe/MgO/Fe due to the unbalance of interface states.[4,21] Such incomplete Fe/MgO interfaces disturb the ideal spin-polarized current transport process, leading to the suppression of TMR ratios in experiments. Therefore, improving the Fe/MgO interface structure through careful process optimization can result in more symmetric *I–V* characteristics. It is one of the necessary conditions for obtaining a giant TMR ratio comparable to the theoretical values.

In this letter, we report large TMR ratios in epitaxial Fe/MgO/Fe(001) exchange-biased spin-valve MTJs up to 417% at RT and 914% at 3 K by careful tuning of interface states using magnetron sputtering, electron-beam (EB) evaporation, post-annealing, and natural oxidation. Inserting ultrathin CoFe at the bottom of the MgO interface further improves the RT-TMR ratio to 496%. We also observed significant TMR oscillations with a very large peak-to-valley difference of ~80% as a function of the MgO thickness. Symmetric differential conductance (*dI/dV*) spectra with fine structures were found for the parallel magnetic configuration, which may be the origin of the highly spin-polarized tunneling current owing to optimum electronic states at the Fe/MgO interfaces.

Exchange-biased spin-valve MTJ multilayers were fabricated with a magnetron sputtering apparatus (ULVAC, Inc.) at a base pressure of $4 \times 10^{-7}$ Pa combined with an EB evaporator. Multilayers were deposited on a 2×2 cm$^2$ single-crystal MgO(001) substrate using the following stack structures [see Fig. 1 (a)]: MgO substrate/Cr (60)/Fe (30 or 50)/Mg (0.5)/wedge-shaped MgO ($d_{MgO}$ = 0.5–2.5)/natural oxidation/Fe (5)/$Ir_{20}Mn_{80}$ (IrMn) (10)/Ru (20) (thickness in nm). The MgO substrate was *in-situ* annealed at 800°C to remove surface contaminations before deposition. All metallic layers were deposited by DC magnetron sputtering at RT and subsequently *in-situ* post-annealed in the sputtering chamber to improve flatness and crystallinity of each layer. The Mg layer was inserted to protect oxidation during the MgO deposition.[22] The MgO barrier was deposited at RT using EB-evaporation of a sintered MgO plate (Kojundo Chemical Laboratory Co., LTD.) with a back pressure $P_{depo}$ of ~1 × $10^{-5}$ Pa and deposition rate of 8×$10^{-3}$ nm/s. For the wedge-shaped MgO barrier, we used a linear motion shutter between the MgO plate and the substrate. We prepared two samples with different Cr buffer conditions: no-annealing or post-annealed at 600°C. The bottom- (top-) Fe was annealed at 300°C (400°C). The barrier was post-annealed at 250°C followed by 300 s natural oxidation using pure $O_2$ gas (99.999%, ~5 Pa) after cooling to RT to control the MgO/top-Fe interface condition. An additional MTJ with a CoFe insertion, i.e., substrate/Cr (60)/Fe (50)/$Co_{50}Fe_{50}$ (2.24)/Mg (0.5)/MgO (2.15)/natural oxidation/Fe (5)/IrMn (10)/Ru (20) (thickness in nm), was also prepared to tune the bottom-Fe interface state. The complete stack was *ex-situ* annealed in a 0.7 T magnetic field at 200°C along with the MgO[110]||Fe[100] direction. Unpatterned films were characterized using an atomic force microscope (AFM) and 4-axis x-ray diffraction (XRD) with Cu $K\alpha$ radiation (wavelength: 0.15418 nm) together with a graphite monochromator. A TMR ratio and a resistance-area product (*RA*) of unpatterened films were measured using the current in-plane tunneling (CIPT) (Capres A/S,

CIPTech-SPM200 prober).[23]

Multilayer stacks were patterned into 6–39 μm$^2$ area elliptical junctions with the long axis parallel to the Fe[100] easy axis using EB-lithography, photolithography, and Ar-ion etching. The microfabricated MTJs were evaluated using a conventional DC 4-probe method at RT using a sourcemeter (Keihley, 2400) and nanovoltmeter (Keithley, 2182A). The TMR ratios of microfabricated MTJs were carefully measured with a constant current to maintain <10 mV bias voltage and reverse the bias polarity for each measurement point to cancel the thermal electromotive force. The device resistance range of prepared MTJs was between 1.0 Ω and 2.1×10$^4$ Ω. A physical property measurement system (PPMS) (Quantum Design, Dynacool) apparatus was used to characterize the temperature dependence of the TMR ratio and resistances from RT to 3 K. In this study, a negative bias voltage corresponds to electrons tunneling from the bottom to the top electrode. The TMR ratio is defined as $(R_{AP}–R_P)/R_P \times 100\%$, where $R_{[AP]P}$ is the resistance in the [antiparallel (AP)] parallel (P) magnetization configuration. High-resolution high-angle annular dark-field scanning transmission electron microscopy (HAADF-STEM), nano-electron-beam diffraction (NEBD), and energy dispersive X-ray spectrometer (EDS) (Titan G2 80-200 TEM) were used to investigate the microstructure of the MTJ cross-section.

Figure 1 (b) and (c) show RT-TMR ratio and $RA$ as a function of $d_{MgO}$ for three different MTJs without Cr annealing (Wafer-1, blue), 600°C Cr post-annealing with 30 nm Fe (Wafer-2, black), and Cr post-annealing with 50 nm Fe (Wafer-3, green). The thickness of the wedge-shaped barrier was calibrated by STEM images taken at two different wafer positions along with the barrier wedge of Wafer-2. The TMR ratios increases rapidly for $d_{MgO} > 1$ nm reaching 400% for MTJs with the Cr post-annealing. The value of Wafer-1 with no-annealed Cr remains at around 280%. The TMR ratio oscillates with $d_{MgO}$ for all displayed MTJs; especially, a maximum peak-to-valley difference of ~80% was observed in Wafer-2 and Wafer-3, is approximately seven times larger than the earlier report by Yuasa *et al.* for Fe/MgO/Fe.[4] The oscillation period is ~0.3 nm, consistent with their work.[4,24] All the prepared wafers were matched to the calibrated Wafer-2 by the oscillation peak position and *RA*. The latter shows an exponential increase with $d_{MgO}$ above 1 nm, suggesting a well-defined MgO thickness and good MTJ reproducibility. Also, oscillating components of *RA* seem to be present as reported.[4,24] The TMR ratios at their peak are 283% at $d_{MgO} = 1.55$ nm for Wafer-1, 405% at $d_{MgO} = 1.50$ nm for Wafer-2, and 417% at $d_{MgO} = 2.16$ nm for Wafer-3, respectively. The TMR ratio of Wafer-3 is slightly larger than that of Wafer-2, presumably owing to an improved bottom Fe quality; however, the main features, such as peak TMR ratio, oscillation peak positions, and the *RA* values are almost identical. Figure 1 (d) shows the TMR-magnetic field (*H*) curve for the Wafer-3 MTJ ($d_{MgO} = 2.16$ nm) with the maximum TMR ratio of 417%. It is much larger than the reported values in pure-Fe-based electrode MTJs, e.g., epitaxial Fe/MgO/Fe(001) ~219%,[25] epitaxial Fe/MgAl$_2$O$_4$/Fe(001) ~248%,[26] and textured FeB/MgO/FeB ~300%.[7]

The star in Figs. 1 (b) and (c) indicates an MTJ with a 16 ML (1 ML = 0.14 nm) CoFe insertion at the bottom interface with a 2.16 nm MgO and Cr post-annealing condition. The CoFe-inserted MTJ shows a further increased TMR ratio of 496%, as seen in its TMR-*H* curve [Fig. 1 (e)] owing to improved bottom-MgO side interface and increased effective tunneling spin polarization by CoFe.[27]

To verify the observed large TMR ratios, we evaluate the unpatterned Wafer-2 using CIPT (see supplementary Fig. S1). The TMR ratios exceeded 400% and TMR oscillations with the 0.3 nm period were well reproduced in a wide *RA* range by the CIPT fitting. It is worth knowing that the TMR oscillation's possible origin is a longstanding question, although many theoretical calculations were performed, especially on the 0.3 nm period's universality.[2,28,29] Further systematic studies using MTJs with such a large oscillation amplitude may further discuss the possible origin.

Figure 2 shows the bias voltage dependencies of the (a) AP and (b) P normalized conductance ($G = dI/dV$) at RT using DC 4-probe method for Wafer-1 and Wafer-2 with two different MgO thicknesses. The conductance was numerically calculated from *I-V* curves. They appear almost symmetric, indicating similar bottom- and top-Fe/MgO interfaces (and their electronic structures) nearly independent of the Cr annealing. In previous reports, stronger asymmetry was observed.[4,30,31] It indicates a difference in the crystallinity of the bottom- and top-Fe/MgO interfaces due to over-oxidation, under-oxidation, or density of misfit dislocations.[19,32–35] Such imperfections can alter the electronic structures of the respective interfaces, reducing TMR significantly.[4,20,21] Therefore, we expect well-tuned bottom- and top-Fe/MgO interfaces were achieved through fine-tuning of interface conditions, such as the Mg insertion, post-annealing temperature, and post-oxidation conditions, resulting in nearly doubled RT-TMR ratios in our MTJs. The P conductance exhibits two wide and deep minima from −0.58 to −0.28 V and from +0.30 V to +0.57 V, respectively. The curves of both samples are similar in shape. The main difference between the Wafer-1 and Wafer-2 is observed in the AP state in which the latter curve shows steeper slopes.

To understand the possible origin of the difference between the Wafer-1 and Wafer-2 despite the similar $dI/dV$ features, we evaluated the degree of the (001)-orientation and film flatness of the bottom-Fe structure using XRD and AFM, respectively. Figure 3 (a) shows the out-of-plane $2\theta$-$\omega$ XRD profiles of stacks with MgO(001) substrate//Cr (60 nm)/Fe (30 nm)/Ru cap (2 nm) with as-deposited (red) and 600°C annealed Cr (black). Growth with a (001)-orientation was confirmed for both samples; however, the full width at half maximum (FWHM) of the Fe(002) rocking curves ($\omega$ scan) of the annealed Cr sample was narrower than that of the as-deposited, as shown in the inset of Fig. 3 (a). In Figs. 3 (b) and (c), the AFM images show a smooth surface without defects for both the samples: peak-valley ($P$-$V$) difference of 1.25 nm (1.10 nm) and average roughness $R_a$ = 0.11 nm (0.10 nm) for the as-deposited (annealed Cr) sample. The annealed Cr sample showed well-developed terrace structures, which may be related to the improved (001)-orientation. These results suggest that the improved degree of (001)-orientation and achievement of a flat bottom-Fe surface contributed to the TMR enhancement due to the suppression of spin-scattering processes.

Figure 4 shows the temperature dependences of the (a) TMR ratio, (b) $R_P$, and $R_{AP}$ for an Fe/MgO/Fe MTJ with a TMR ratio of 386% at 300 K (Wafer-2 with $d_{MgO}$ = 2.18 nm). The low temperature (LT) and RT-TMR-$H$ curves are shown in the upper panel's inset. The TMR ratio increases with reduced temperature, reaching 914% at 3 K. This LT value is much larger than typical Fe/MgO/Fe (250~370%).[4,24,25,31] We obtained the effective tunneling spin polarization at LT ($P_0$) to be 0.905 using the Jullière equation: TMR = $2P_0^2/(1 - P_0^2)$,[36] suggesting close to half-metallic $\Delta_1$ states in our Fe/MgO/Fe structure. The temperature dependence of the TMR ratio is mainly determined by that of $R_{AP}$, which is the typical behavior in MgO and MgAl$_2$O$_4$ based MTJs.[26,31]

Figure 4 shows the bias voltage dependences of (c) $G_{AP}$ and (d) $G_P$ at RT and 3 K. Fine structures could be observed only in the $G_P$ spectra, indicated by arrows and labeled A~C. At 3 K, two narrower minima at negative and positive bias voltage appeared. The minima B at ±0.28 V observed at RT remain almost unchanged at 3 K. Additional minima C appeared at ±0.55 V and shoulders A at ±0.08 V. The B minima are typically observed in MTJs exhibiting large TMR ratios.[20,37,38] To the authors' knowledge, the C minima which appeared at LT have not been reported in MgO-based MTJs. Considering the Fe bulk electronic structure, no energy states are expected around 0.55 eV from the Fermi energy. Therefore, we associate the observed minima with Fe/MgO interface electronic states. At LT, the $G_{AP}$ showed a substantial dip at the zero-bias, indicating that magnon-assisted tunneling dominates the transport,[39] consistent with previous Fe/MgO/Fe and Fe/MgAl$_2$O$_4$/Fe MTJs.[26,40] According to the $G_P$ and $G_{AP}$ spectra, the large TMR ratio at a low bias voltage and LT is attributed to enhancing the zero-bias conductance for the P state and the suppressing the zero-bias conductance for the AP state.

Figure 5 (a) shows the cross-sectional HAADF-STEM image of the Fe/MgO/Fe (Wafer-2). The image was taken at an MgO thickness of $d_{MgO}$ = 2.3 nm. Relatively flat Fe/MgO interfaces are observed; however, no qualitative difference to other reports on Fe/MgO/Fe MTJs with a smaller TMR ratio,[4,41] which is in contrast to Fe/MgAl$_2$O$_4$/Fe MTJs with misfit-free interfaces.[26] The number of misfit dislocations could be observed from the fast Fourier transform (FFT) filtered image of Fig. 5 (a) as displayed in Fig. 5 (b). Both interfaces have misfit dislocations commonly observed in Fe/MgO/Fe due to the 3% mismatch. NEBD patterns of the bottom-Fe electrode, MgO barrier, and top-Fe electrode (supplementary Fig. S2) showed that all the layers grow epitaxially with an epitaxial relationship of Fe[110](001)∥MgO[100](001)∥Fe[110](001) as expected. EDS elemental profiles of the observed area are shown in Fig. 5 (c). It is worth knowing that the O signal outside the barrier is an artifact originating from the TEM specimen's surface. The profiles show elementally sharp MgO interfaces and no significant atomic diffusion from the upper electrode side. Therefore, the MgO interface flatness was maintained even after depositing the MgO barrier and upper electrode. Further TMR ratio improvement can be expected if the misfit dislocation density can be reduced by maintaining the interface sharpness.

Microstructural changes at the bottom- and top-Fe/MgO interfaces are beyond our STEM observation's resolution limit. The symmetric conductance spectra and their specific local minimum structures show improved interface states; sharp interfaces without oxidization of the Fe surfaces may have been achieved, creating close to ideal interfaces at both barrier sides. Our results demonstrated enough room for even larger TMR ratios in various MTJs by developing an advanced interface structure process.

In summary, giant RT TMR ratios up to 417% (914% at 3K) in Fe/MgO/Fe and 496% in Fe/CoFe/MgO/Fe exchange-biased spin-value MTJs were obtained by fine-tuning of MgO interface states and improved (001)-orientation. In accordance with the improved TMR ratio, significant TMR oscillations exceeding 80% with respect to the MgO thickness was observed at RT. Symmetric differential conductance spectra with remarkable fine structures were observed in the parallel state. These results suggested that the interface modification effectively improved the electronic states of the MgO/Fe interfaces, enabling significant enhancement in TMR ratios even in Fe/MgO/Fe. This study suggests that much larger TMR ratios can be achieved by further improving the film process and material technologies with a greater focus on the interface structures, which will open up the possibility of future spintronic applications.

**SUPPLEMENTARY MATERIALS**

See supplementary material for the CIPT measurement and NEBD refraction patterns.

**ACKNOWLEDGEMENT**

The authors would like to thank H. Ikeda for technical support, J. Uzuhashi for TEM characterization, Y. Miura and K. Masuda for their valuable comments. This work was partly supported by the ImPACT Program of the Council for Science, Technology and innovation (Cabinet Office,

Government of Japan), JSPS KAKENHI Grant No. 16H06332, and TIA collaborative research program "Kakehashi". This paper is partly based on results obtained from a project, JPNP16007, commissioned by the New Energy and Industrial Technology Development Organization (NEDO).

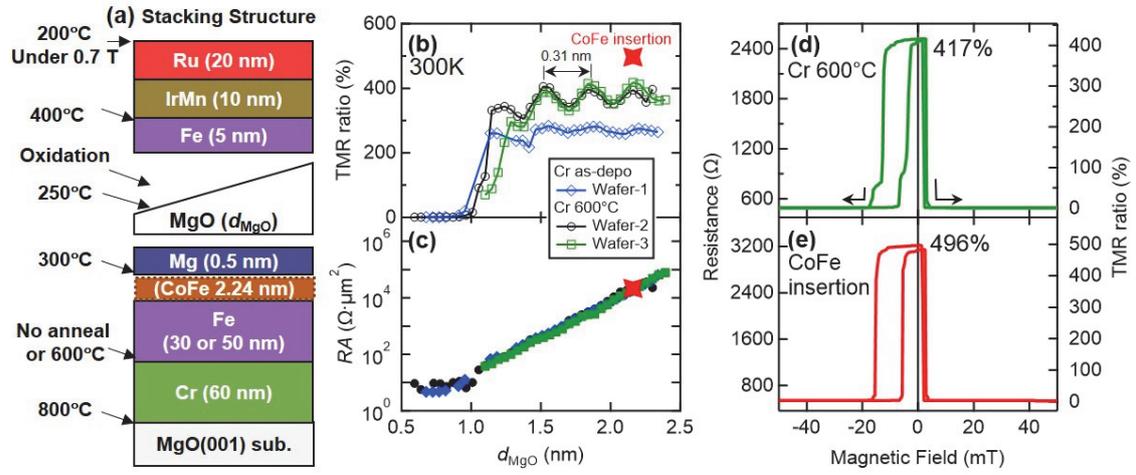

**FIG. 1.** (a) Schematic MTJ stacking structure. (b) TMR ratio and (c) $RA$ vs. $d_{MgO}$ evaluated by DC 4-probe method for Fe/MgO/Fe (Wafer-1, 2, and 3), and Fe/CoFe (2.24 nm)/MgO/Fe. (d) and (e) TMR-$H$ curves at RT for the (c) Wafer-3, and (d) Fe/CoFe/MgO/Fe MTJs.

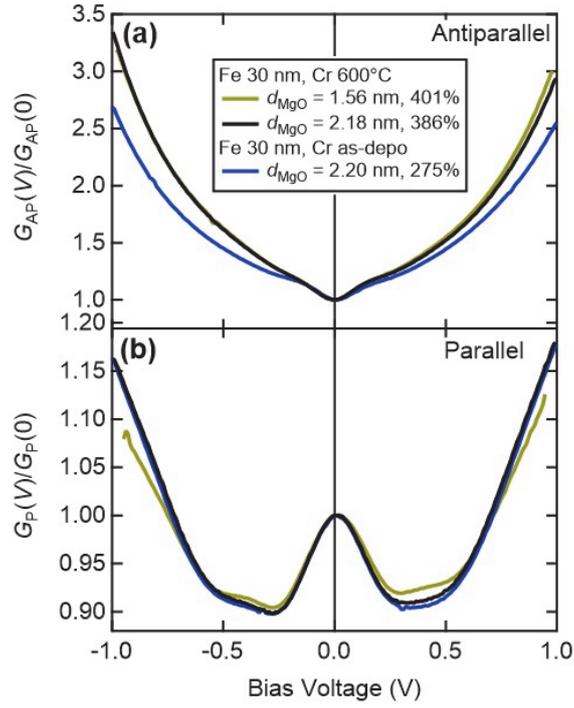

**FIG. 2.** Normalized conductance ($G$) spectra for (a) AP and (b) P state at RT of MTJs with a 30 nm bottom Fe electrode. Results of as-deposited Cr with $d_{MgO}$ = 2.20 nm (Wafer-1), and Cr-annealed with $d_{MgO}$ = 1.56 and 2.18 nm (Wafer-2) are shown.

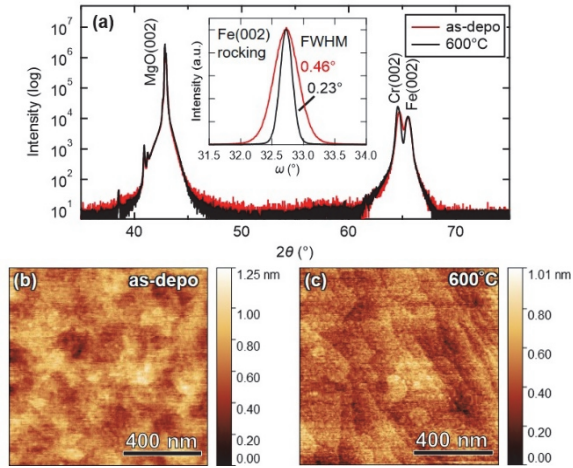

**FIG. 3.** (a) Out-of-plane XRD scans of MgO substrate//Cr(60 nm)/Fe(30 nm)/Ru(2 nm) with as-deposited and 600°C annealed Cr-buffer. Inset of (a) indicates respective Fe rocking curves. (b) and (c) 1×1 μm² area AFM images of (b) as-deposited and (c) 600°C annealed Cr-buffer samples.

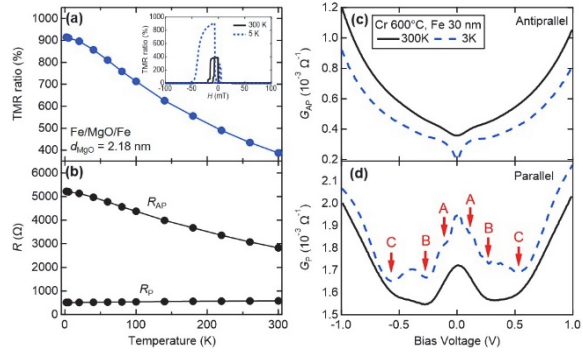

**FIG. 4.** (a) Temperature dependence of (a) TMR ratio and (b) $R_P$ and $R_{AP}$ for MTJ with $d_{MgO}$ = 2.18 nm (Wafer-2). Inset of (a): TMR-$H$ curves at 300 K (solid, black) and 5 K (dashed, blue). (c) and (d) $G$ spectra at 3 K and RT for (c) AP and (d) P states.

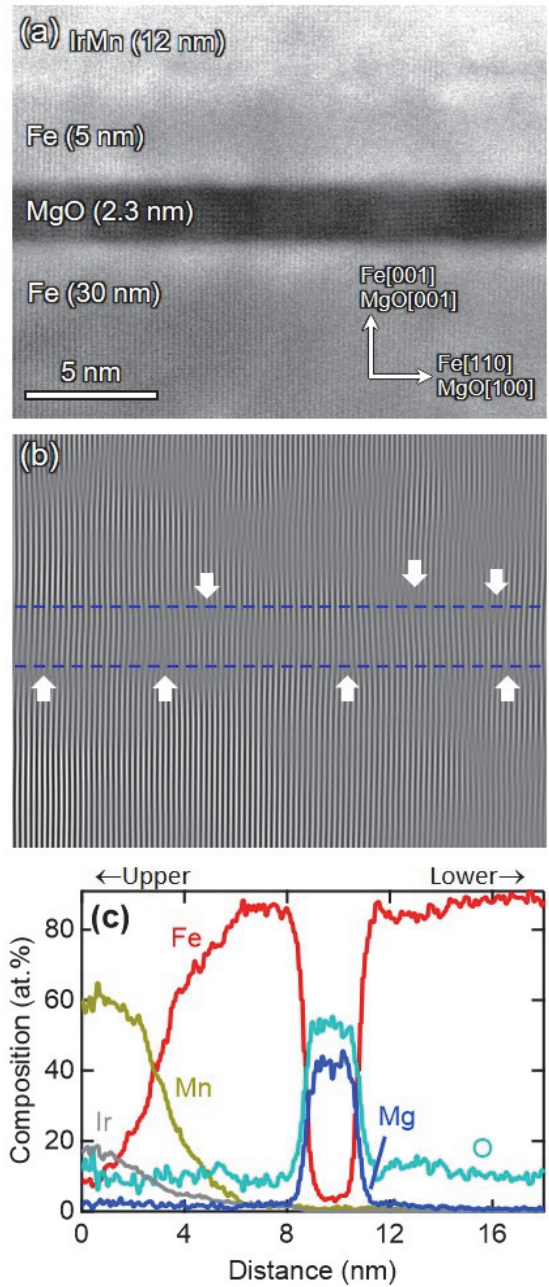

**FIG. 5.** (a) Cross-sectional HAADF-STEM image of Wafer-2 with $d_{MgO}$ = 2.3 nm. (b) FFT filter images using (a). Arrows in (b) indicate lattice dislocations at the MgO interfaces. (c) Elemental depth profiles using EDS.

Supplementary Material

**1. CIPT Measurement**

In this study, CIPT measurements were performed using the wide type M12PP Micro Twelve-Point Probes (Capres A/S, M12PP_005), which can evaluate an MgO thickness $d_{MgO}$ range between 1.1 and 2.2 nm with reliable fitting based on the theoretical equations.[1] The probe has a minimum (maximum) mean probe spacing of 3.0 (59.0) μm.

Results of current in-plane tunneling (CIPT) measurements of the unpatterned Wafer-2 (600°C Cr post-annealing with 30 nm Fe) are summarized in Fig. S1. Figures S1 (a) and (b) show the tunnel magnetoresistance (TMR) ratio and resistance area (RA) product as a function of an MgO thickness $d_{MgO}$, respectively. The results of the patterned MTJ data for the same Wafer-2 was also plotted in Figs. S1 (a) and (b). Figure S1 (c) shows the sheet resistances of the metallic layer above and below the MgO barrier, $R_t$ and $R_b$, respectively. The TMR ratio, RA, $R_t$, and $R_b$ were obtained as the CIPT fitting parameters. The TMR ratio and RA of the CIPT data points were consistent with the values of the patterned MTJ ones; the large TMR ratios exceeding 400% were observed even by the CIPT measurement of the unpatterned wafer. Surprisingly, the significant TMR oscillation with the ~0.3 nm period was also clearly observed by CIPT, indicating that the oscillation feature does not depend on an MTJ area and patterning processes. Additionally, as seen in Fig. S1 (c), constant $R_t$ and $R_b$ are observed in a wide range of RA, which ensures the good fitting quality. Close to $d_{MgO}$ = 1.1 nm and 2.2 nm some deviations of $R_t$ and $R_b$ can be observed, indicating the edges of the reliable fitting range for the used probe. In Fig. S1 (d), a typical CIPT fitting result is displayed. $R_{sq}^{low}$ (lower panel) is the measured sheet resistance per square in the parallel (P) magnetization state, and TMR$_{cip}$ (upper panel) is the current-in-plane TMR ratio. The measured data were well reproduced by the model. The fit results (TMR ratio, RA, $R_t$, and $R_b$) are displayed in the box of the lower panel.

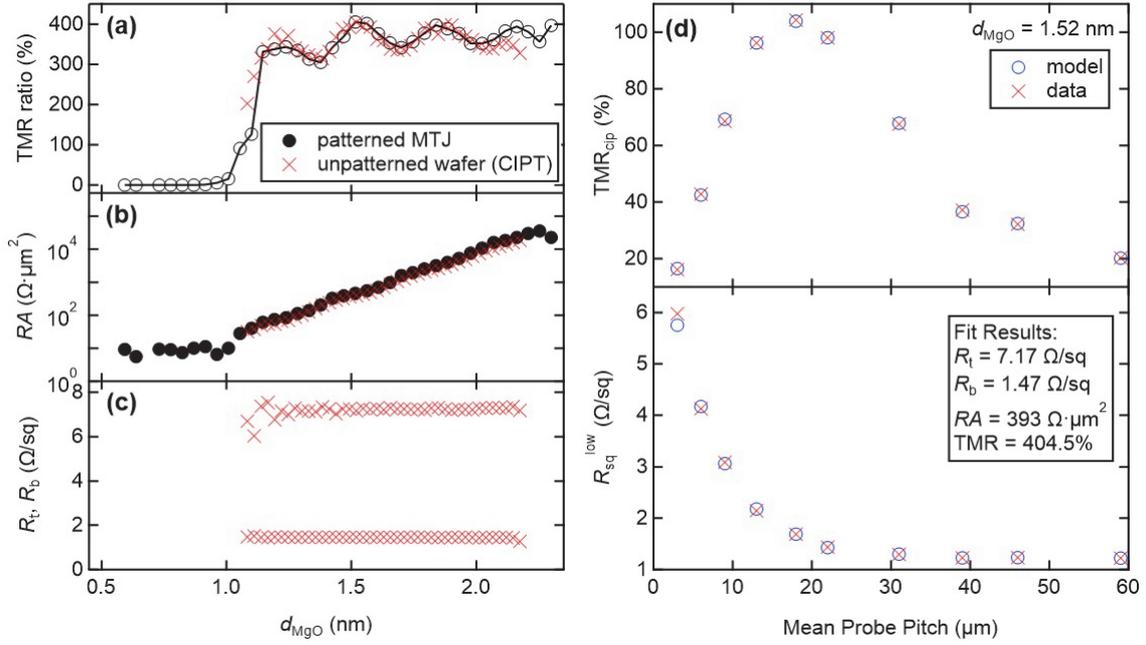

**FIG. S1.** (a)-(c) Summary of CIPT results of Wafer-2 at room temperature: (a) TMR ratio, (b) $RA$, (c) $R_t$ and (d) $R_b$. For (a) and (b), patterned MTJ results were also displayed for comparison. (d) CIPT fitting results at $d_{MgO}$ = 1.52 nm; upper (lower) panel is current-in-plane TMR ratio (sheet resistance per square in P state) data fitting by the theoretical model.

## 2. Nano Electron-Beam Diffraction Patterns

Fig. S2 displays the nano electron-beam diffraction (NEBD) patterns of the top-Fe, MgO barrier, and bottom-Fe. The layers have a cubic structure with Fe being bcc and MgO rock-salt.

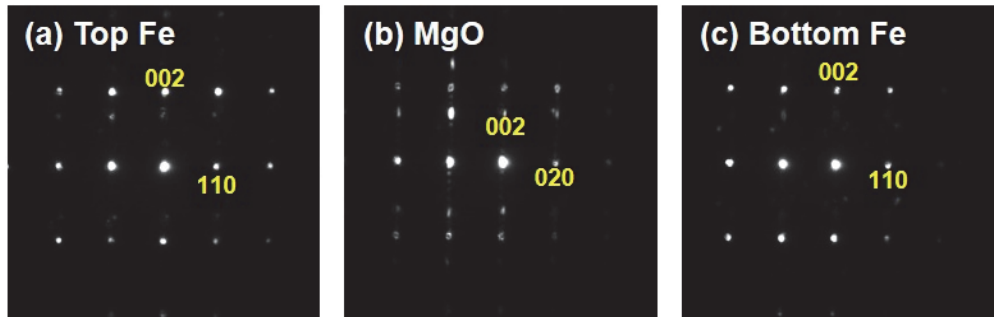

**FIG. S2.** NEBD patterns for (a) top-Fe, (b) MgO barrier, and (c) bottom-Fe; reflection indices are labeled.

**Reference**
[1] D.C. Worledge and P.L. Trouilloud, Appl. Phys. Lett. **83**, 84 (2003).